\documentclass[a4paper]{article}

\usepackage{INTERSPEECH2022}
\usepackage{url}

\title{The TEA-ASLP System for Multilingual Conversational Speech Recognition and Speech Diarization in MLC-SLM 2025 Challenge}
\name{Hongfei Xue$^{1,2}$, Kaixun Huang$^1$, Zhikai Zhou$^1$, Shen Huang$^1$, Shidong Shang$^1$}
\address{
  $^1$Tencent Ethereal Audio Lab, Tencent Corporation, Beijing, China \\
  $^2$Audio, Speech and Language Processing Group (ASLP@NPU)
  }
\email{\{hongfeixue, kaixunhuang, adrenzhou, springhuang\}@tencent.com}
  
\begin{document}

\maketitle
\begin{abstract}
This paper presents the TEA-ASLP's system submitted to the MLC-SLM 2025 Challenge, addressing multilingual conversational automatic speech recognition (ASR) in Task I and speech diarization ASR in Task II. For Task I, we enhance Ideal-LLM model by integrating known language identification and a multilingual MOE LoRA structure, along with using CTC-predicted tokens as prompts to improve autoregressive generation. The model is trained on approximately 180k hours of multilingual ASR data. In Task II, we replace the baseline English-Chinese speaker diarization model with a more suitable English-only version. Our approach achieves a 30.8\% reduction in word error rate (WER) compared to the baseline speech language model, resulting in a final WER of 9.60\% in Task I and a time-constrained minimum-permutation WER of 17.49\% in Task II, earning first and second place in the respective challenge tasks.
\end{abstract}

\section{Introduction}
\label{sec:intro}
Text-based Large Language Models (LLMs) have had a profound impact on the field of artificial intelligence, thanks to their advanced capabilities in understanding and generating natural language~\cite{openai2022chatgpt, openai2023gpt4, brown2020language, anil2023palm, LLaMA}. Recently, there has been growing interest in combining LLMs with audio encoders, enabling the models to process and understand audio modalities~\cite{gong2023listentu, tang2023salmonn, chu2023qwenaudio, 24wavllm, chu2024qwen2audio, 24echat}.
A key component in audio processing is multilingual automatic speech recognition (ASR), a challenging task that has been enhanced by integrating speech encoders with LLMs via connectors. This approach has been shown to outperform conventional end-to-end models in terms of performance~\cite{meta24llmasr, 23speechllama, chu2024qwen2audio, 24slam-asr, bai2024seed}.
For instance, a method described in~\cite{meta24llmasr} employs a connectionist temporal classification (CTC) trained encoder to handle speech sequences, which are then passed through a projection layer to an LLM decoder. 
Recent advancements in multilingual ASR have also been driven by the use of Whisper encoders~\cite{23whisper} and self-supervised learning (SSL) encoders, which have led to substantial improvements~\cite{chu2023qwenaudio, 24slam-asr, bai2024seed}. The Qwen2-Audio model~\cite{chu2024qwen2audio} leverages a fine-tuned Whisper encoder~\cite{23whisper} for speech feature extraction, resulting in notable improvements in multilingual ASR tasks.


However, the advancement of robust LLM-based spoken dialogue models is highly dependent on real-world conversational speech data, which captures the complexities of human communication, such as natural pauses, interruptions, speaker overlaps, and varied conversational styles. The limited availability of such data, especially in multilingual settings, presents a significant obstacle to further progress in this area.
Inspired by this, a workshop has been proposed for Interspeech 2025, aiming to bridge the gap by hosting a challenge to build multilingual conversational speech language models (MLC-SLM) alongside the release of a real-world multilingual conversational speech dataset.

This study presents our system for the multilingual ASR (Task I) and speech diarization ASR (Task II) in the MLC-SLM 2025 Challenge. For Task I, we build on our previous work, Ideal-LLM~\cite{xue2024ideal}, which utilizes dual multilingual encoders~\cite{23whisper, 23mms} to enhance language representations and a language-specific connector for language adaptation. Since language identification (LID) is known in this task, we further enhance it with a multilingual MoE LoRA~\cite{22lora} (mLoRA) adapter, routed by LID. Additionally, we leverage CTC-predicted tokens from the connector as non-autoregressive outputs to support autoregressive generation. For data, we use approximately 180k hours of multilingual ASR data, applying a robust ASR model to filter out erroneous samples.
For Task II, we adopt a pipeline-based approach, combining a 3D-Speaker Diarization model with the pre-trained Task I ASR model. The baseline SD model, initially designed for both English and Chinese, is replaced with a more suitable English-only model.
Experimental results show that our model is more effective at distinguishing languages and aligning the multilingual embedding space. Specifically, our approach yields a substantial boost in ASR performance, achieving a 30.8\% relative reduction in average word error rates (WER) compared to the Whisper encoder integrated with LLMs when using only the MLC-SLM data. After incorporating all available data, our system achieved a WER of 9.60\% and a time-constrained minimum-permutation word error rate (tcpWER) of 17.49\% on the two evaluation sets, earning first and second places in the respective challenge tasks.

\begin{figure*}[t]
\centering
\includegraphics[width=0.9\linewidth]{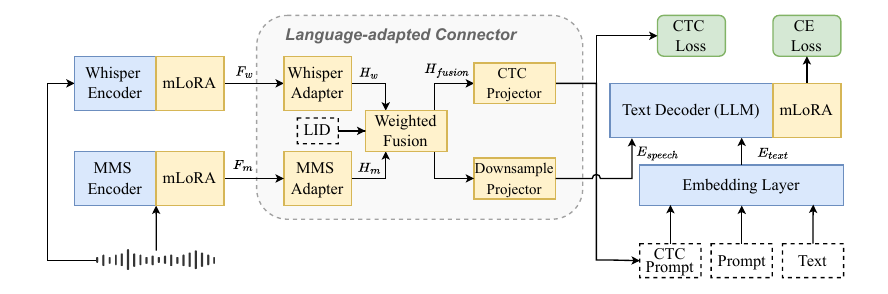}
\caption{
The overall framework of the TEA-ASLP system.}
\label{fig:model}
\vspace{-15pt}
\end{figure*}

\section{Proposed System}

\label{sec:proposed_method}
\subsection{Architecture}

We adopt the Ideal-LLM structure, which includes dual encoders, a language-adapted connector, and a text decoder. To more effectively leverage the known LID information, we replace the original LLM LoRA~\cite{22lora} adapter with a multilingual MoE LoRA (mLoRA) adapter, routed according to the specific LID. Additionally, we incorporate CTC prompts to assist in the LLM generation process. An illustration of the overall architecture is shown in Fig.~\ref{fig:model}.

\noindent\textbf{Dual Encoders} 
Our dual speech encoders are based on Whisper~\cite{23whisper} and MMS~\cite{23mms}, both of which are robust models trained on large multilingual datasets using weakly-supervised and self-supervised learning, respectively. The representations from these models complement each other due to their distinct pre-training methods on diverse language distributions. The speech signal is fed into both the Whisper and MMS encoders to generate the speech features $F_w$ and $F_m$.

\noindent\textbf{Language-adapted Connector} 
Since the dual encoders have been trained on different language distributions, we design a connector to perform a language-dependent fusion of the dual encoders' features, transforming them into the embedding space of the LLM. First, the speech features $F_w$ and $F_m$ are transformed into hidden representations by Whisper and MMS adapter, which are transformer encoder networks~\cite{17attention}, resulting in $H_w$ and $H_m$. Then, based on the weighted fusion module, $H_w$ and $H_m$ are mixed with different weights to form $H_{fusion}$. 
The fused hidden representations $H_{fusion}$ are input into a CTC Projector to obtain the CTC token, which serves as a context prompt for the embedding layer.
Additionally, $H_{fusion}$ is sequentially down-sampled through the convolution layer in the Downsample Projector, and the projection layer maps it to $E_{speech}$ in the LLM embedding space.

In the weighted fusion module, we initialize trainable parameters for each language and apply a sigmoid function to generate weights. These weights is selected by the known LID for fusion. This process is learned through backpropagation with decoder loss and CTC loss, guiding the model to prefer certain encoders based on language-specific features.

\noindent\textbf{Text Decoder} 
The text decoder is based on the Qwen-3-8B base model~\footnote{https://huggingface.co/Qwen/Qwen3-8B-Base}, a language model with 8 billion parameters trained on 36 trillion tokens~\cite{yang2025qwen3}.
The prompts and text labels are represented by the tokenizer embedding layer, which is then concatenated with $E_{speech}$ from the language-adapted connector. The introduction of CTC Prompt helps mitigate the hallucinations of the LLM, thereby reducing insertion errors.
These embeddings are subsequently passed into the text decoder, with the target output being the text labels.

\noindent\textbf{mLoRA} 
Given the inherent differences between languages, we propose a multilingual MoE LoRA adapter for training. Each language is assigned a specific LoRA adapter, and the speech features are routed to the corresponding adapter based on the known LID. The mLoRA adapter is applied to both the dual encoders and the text decoder.

\subsection{Multi-stage Training}
We adopt a two-stage training strategy to improve the multilingual speech LLM’s ASR capability.

\noindent\textbf{Stage 1}
In the first stage, we train the dual encoders and the language-adapted connector using only CTC loss. The CTC token labels are encoded using the same LLM tokenizer to align $H_{fusion}$ with the LLM embedding space. This training process is performed in two steps: initially, both encoders are fully unfrozen; in the second step, the encoders are frozen, and mLoRA training is introduced.

\noindent\textbf{Stage 2}
In the second stage, we train the downsample projector and text decoder LoRA using only cross-entropy (CE) loss. After the large-scale CTC training in Stage 1, $H_{fusion}$ has become a text-relevant hidden representation, so the earlier modules do not require further training. In these two stages, we perform large-scale pretraining using the full dataset and fine-tune with the MLC-SLM data.

\subsection{SD ASR}
We employ a pipeline that first splits the audio files using a VAD model and then clusters the speakers' embeddings from a speaker verification model. The split audio segments are subsequently fed into the robust ASR model from Task I.
For the SD model, we replace the original English-Chinese speaker verification model with a more robust English-only speaker verification model\footnote{modelscope.cn/models/iic/speech\_eres2net\_large\_sv\_en\_voxceleb\_16k}.
Additionally, we notice that the segmentation results from the baseline model often contained several consecutive audio segments from the same speaker. To address this, we concatenate these adjacent segments, creating longer speech segments that provide more context for improved recognition.

\begin{table}[]
\centering
\caption{Sources of the training datasets.}
\begin{tabular}{@{}lcc@{}}
\toprule
Data Source               & Language    & Duration (kh) \\ \midrule
MLC-SLM                   & 11          & 1.5           \\
MSR-86K~\cite{MSR-86K}                   & 10          & 63.7          \\
CommonVoice~\cite{19commonvoice}               & 11          & 6.0             \\
Multilingual Librispeech~\cite{20mllibrispeech}  & 6           & 49.3          \\
GigaSpeech2~\cite{yang2024gigaspeech}               & 2           & 16.3          \\
Emilia~\cite{Emilia24}                    & 4           & 26.3          \\
OpenDataLab~\cite{he2024opendatalabempoweringgeneralartificial}               & 4           & 0.8           \\
Librispeech~\cite{Librispeech15}, Gigaspeech~\cite{chen2021gigaspeech} & \textit{en} & 3.5             \\
fisher~\cite{cieri-etal-2004-fisher}, swbd & \textit{en} & 2.2             \\
Reazonspeech~\cite{fujimoto2016reazonspeech}, LaboroTV~\cite{ando2021construction}              & \textit{ja} & 7.0             \\
Golos~\cite{karpov2021golos}                     & \textit{ru} & 1.2           \\
Ksponspeech~\cite{bang2020ksponspeech}               & ko          & 1.0             \\
\textbf{Sum}                       & \textbf{11}          & \textbf{179}         \\ \bottomrule
\end{tabular}
\label{tab: dataset}
\vspace{-10pt}
\end{table}

\begin{table*}[]
\centering
\caption{WER (\%) and CER (\%) results on the MLC-SLM development set for various methods.}
\begin{tabular}{@{}lcccccccccccc@{}}
\toprule
Model                    & \multicolumn{1}{c}{\textit{en}} & \multicolumn{1}{c}{\textit{fr}} & \multicolumn{1}{c}{\textit{de}} & \multicolumn{1}{c}{\textit{it}} & \multicolumn{1}{c}{\textit{ja}} & \multicolumn{1}{c}{\textit{ko}} & \multicolumn{1}{c}{\textit{pt}} & \multicolumn{1}{c}{\textit{ru}} & \multicolumn{1}{c}{\textit{es}} & \multicolumn{1}{c}{\textit{th}} & \multicolumn{1}{c}{\textit{vi}} & \multicolumn{1}{c}{avg} \\ \midrule
Baseline                 & 12.19                  & 33.95                  & 23.47                  & 34.74                  & 20.77                  & 34.02                  & 18.25                  & 14.31                  & 21.67                  & 21.5                   & 21.49                  & 20.62                   \\
Baseline (ours)       & 12.13                  & 21.73                  & 33.03                  & 21.64                  & 29.4                   & 19.95                  & 30.48                  & 18.49                  & 13.77                  & 19.21                  & 21.93                  & 19.87                   \\
+ Dual Encoder           & 10.45                  & 17.53                  & 24.54                  & 16.76                  & 20.74                  & 14.23                  & 25.12                  & 16.04                  & 11.12                  & 16.8                   & 17.74                  & 15.95                   \\
\ + CTC Loss             & 10.34                  & 17.25                  & 23.23                  & 15.82                  & 19.71                  & 13.94                  & 24.98                  & 15.54                  & 11.07                  & 11.66                  & 17.31                  & 14.52                   \\
\ \ + CTC context        & 10.19                  & 18.39                  & 23.35                  & 16.34                  & 19.1                   & 12.4                   & 24.89                  & 15.61                  & 11.01                  & 11.17                  & 16.96                  & 14.26                   \\
\ \ \ + Data Scaling \& mLoRA         & \textbf{7.91}          & \textbf{13.74}         & \textbf{16.89}         & \textbf{11.78}         & \textbf{13.61}         & \textbf{8.64}           & \textbf{19.61}         & \textbf{12.55}        & \textbf{8.37}          & \textbf{8.45}          & \textbf{11.45}         & \textbf{10.62}          \\ \bottomrule
\end{tabular}
\end{table*}

\section{Experiments}
\subsection{Datasets}
We use a large corpus to train our model, totaling 180k hours, as shown in Table~\ref{tab: dataset}. This corpus includes data from 11 languages: English (\textit{en}), French (\textit{fr}), German (\textit{de}), Italian (\textit{it}), Japanese (\textit{ja}), Korean (\textit{ko}), Portuguese (\textit{pt}), Russian (\textit{ru}), Spanish (\textit{es}), Thai (\textit{th}), and Vietnamese (\textit{vi}). We apply a data balancing strategy as described in~\cite{21XLSR}. For the YouTube data, we utilize a ASR model OWSM-CTC~\cite{peng2024owsm} to filter out and remove low-quality speech. Additionally, we apply data augmentation techniques, including spectral enhancement and speed variation during training.

\subsection{Experiment Setup}
For the proposed system, the Whisper encoder is from Whisper Large-v3~\footnote{https://huggingface.co/openai/whisper-large-v3}, and the MMS encoder uses the 1B version~\footnote{https://huggingface.co/facebook/mms-1b}. Both the Whisper Adapter and the MMS Adapter are 2-layer Transformer encoders.
The Adam optimizer is used with a peak learning rate of 2e-4, a warmup period of 2k steps, and 100k training steps for Stage 1. For Stage 2, the adam optimizer has a peak learning rate of 5e-5, a warmup of 2k steps, and 100k training steps. We utilize 32 NVIDIA A100 GPUs, with gradient accumulation to process approximately 200 seconds of data per GPU. For inference, we select the best five models and perform average decoding. For short speech, the recognition is done together with context splicing, followed by result alignment through the alignment algorithm.

\subsection{Experiment Results}

\textbf{ASR Results}
Table 2 presents our ASR results on the MLC-SLM Task I development set. For English (\textit{en}), we report the average WER across five regions. The first row shows the baseline results from the baseline model~\footnote{https://github.com/mubingshen/MLC-SLM-Baseline/tree/main}, which combines a Whisper encoder with LLMs. The configurations for several other models are as follows:

\begin{itemize}
    \item \textbf{Baseline (ours):} This model follows the structure of the baseline model but with the downsampling factor changed to 2x and the LLM LoRA rank increased to 32. The training data includes only the MLC-SLM 1.5k hours dataset. The average WER is reduced by 3.6\% compared to the original baseline.

    \item \textbf{+ Dual Encoder:} In this setup, the original Whisper Encoder is replaced by the Dual Encoder, and the proposed Language-adapted Connector is used. During training, only the Connector and LLM LoRA are trainable. The average WER is reduced by 22.6\% compared to the original baseline.

    \item \textbf{+ CTC Loss:} Based on the Dual Encoder, we further incorporate CTC Loss. Using the two-stage training approach described earlier, we first train the Connector with CTC and then train LoRA with CE. The average WER is reduced by 29.6\% compared to the original baseline.

    \item \textbf{+ CTC Context:} In this configuration, the CTC context prompt is added. The token decoded by the CTC non-autoregressive method is sent to the LLM as part of the prompt. The average WER is reduced by 30.8\% compared to the original baseline.

    \item \textbf{+ mLoRA \& Data Scaling:} The introduction of mLoRA and data scaling is reflected directly in the final results. In addition, compared with other results, this model uses Qwen 3-8B instead of Qwen 2.5-7B. The model follows the two-stage training process described previously. For the LLM’s mLoRA, the rank is set to 64 and the alpha to 32, while for the Encoder’s mLoRA, the rank is set to 32 and the alpha to 16. As a result, the final WER is reduced by 48.4\% compared to the original baseline.

\end{itemize}

\noindent\textbf{SD ASR Results}
Table 3 presents the SD ASR results on the MLC-SLM Task II development set. We replaced the original speaker verification model with ERes2Net-large model, leading to a reduction in speaker error rate. Although there is a certain increase in FA, these multi-detected silences have little impact on the final recognition.
When feeding the original baseline segments files into our ASR model, the TcpWER significantly improved compared to the baseline. Further improvements are achieved after splicing adjacent audio segments, resulting in an additional reduction in TcpWER.

\begin{table}[]
\centering
\caption{MS, FA, and SER on the MLC-SLM dev set for diverse methods, along with TcpWER (\%) results on the test set.}
\begin{tabular}{@{}lllll@{}}
\toprule
Model           & MS   & FA   & SER  & TcpWER \\ \midrule
Baseline        & 1.76 & \textbf{10.25}  & 4.43 & 60.39  \\
Proposed System in Task I & 1.76 & \textbf{10.25}  & 4.43 & 18.56     \\
+ ERes2Net-large \& Concat & \textbf{0.37} & 14.82  & \textbf{3.95} & \textbf{17.49}     \\ \bottomrule
\end{tabular}
\end{table}

\section{Conclusions}

Our system for the MLC-SLM 2025 Challenge demonstrates significant improvements in both multilingual ASR and speech diarization tasks. By enhancing our Ideal-LLM model with language identification and a multilingual LoRA structure, and optimizing the diarization model, we achieved notable reductions in word error rates and secured top positions in the challenge. These results highlight the effectiveness of our approach and contribute to the ongoing advancement of multilingual conversational speech models.


\bibliographystyle{IEEEtran}

\bibliography{mybib}


\end{document}